# Influence of resonances on the $^{11}$B(n,γ)$^{12}$B capture reaction rate. Capture to the ground state of $^{12}$B.

Dubovichenko S.B.[1,2,*], Burkova N.A.[2], Dzhazairov-Kakhramanov A.V.[1,*], Tkachenko A.S.[1,2]

[1]*Fesenkov Astrophysical Institute "NCSRT" ASA MDASI RK, 050020, Almaty, Kazakhstan*
[2]*al-Farabi Kazakh National University, 050040, Almaty, Kazakhstan*

**ABSTRACT:** Within the framework of the modified potential cluster model with a classification of orbital states according to Young diagrams, the possibility of describing experimental data for total cross sections of the neutron radiative capture on $^{11}$B to the ground state of $^{12}$B at energies of 10 meV (1 meV = $10^{-3}$ eV) to 7 MeV was considered. It was shown that, taking into account only the $E$1 transition from the $S$ state of the $n^{11}$B scattering to the ground state of $^{12}$B, it is quite possible to explain the magnitude of the known experimental cross sections at energies of 25.3 meV to 70 keV. Furthermore, on the basis of the total cross sections of 10 meV to 7 MeV, but excluding resonances above 5 MeV, the reaction rate is calculated in the temperature range of 0.01 to 10.0 $T_9$. It is shown that the inclusion of low-lying resonance states makes a significant contribution to the reaction rate, starting already with temperatures of 0.2–0.3 $T_9$.

## 1. Introduction

In recent years, the role of *short-lived isotopes* that are included in various chains of the synthesis of heavier elements in astrophysical processes has been actively discussed. This role is noticeable both in the processes of primordial nucleosynthesis and in nuclear processes during supernova explosions. A well-known yield on the carbon nuclides is proposed as a sequence of processes [1–4]

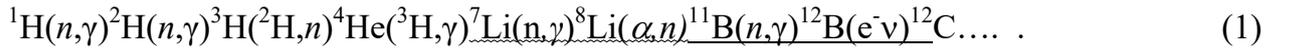

$$^{1}\text{H}(n,\gamma)^{2}\text{H}(n,\gamma)^{3}\text{H}(^{2}\text{H},n)^{4}\text{He}(^{3}\text{H},\gamma)^{7}\text{Li}(n,\gamma)^{8}\text{Li}(\alpha,n)^{11}\text{B}(n,\gamma)^{12}\text{B}(e^{-}\nu)^{12}\text{C}\ldots. \qquad (1)$$

There is another possible option for the developing of this chain, where $^9$Be is existed and what is typical for neutron-excess stars (see, for example, [5])

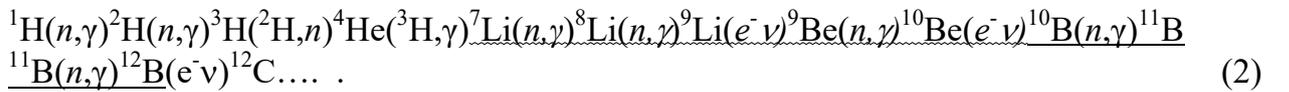

$$^{1}\text{H}(n,\gamma)^{2}\text{H}(n,\gamma)^{3}\text{H}(^{2}\text{H},n)^{4}\text{He}(^{3}\text{H},\gamma)^{7}\text{Li}(n,\gamma)^{8}\text{Li}(n,\gamma)^{9}\text{Li}(e^{-}\nu)^{9}\text{Be}(n,\gamma)^{10}\text{Be}(e^{-}\nu)^{10}\text{B}(n,\gamma)^{11}\text{B}$$
$$^{11}\text{B}(n,\gamma)^{12}\text{B}(e^{-}\nu)^{12}\text{C}\ldots. \qquad (2)$$

However, in some publications [6–8] the chain similar to (1)–(2) is considered, but without $^8$Li as a participant

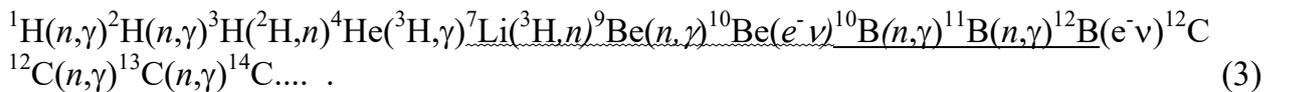

$$^{1}\text{H}(n,\gamma)^{2}\text{H}(n,\gamma)^{3}\text{H}(^{2}\text{H},n)^{4}\text{He}(^{3}\text{H},\gamma)^{7}\text{Li}(^{3}\text{H},n)^{9}\text{Be}(n,\gamma)^{10}\text{Be}(e^{-}\nu)^{10}\text{B}(n,\gamma)^{11}\text{B}(n,\gamma)^{12}\text{B}(e^{-}\nu)^{12}\text{C}$$
$$^{12}\text{C}(n,\gamma)^{13}\text{C}(n,\gamma)^{14}\text{C}\ldots. \qquad (3)$$

Different parts of chains are marked above by wavy line and the boron processes are marked by underlining. Some other scripts of similar processes are given in [9,10].

---
[*] Corresponding authors: albert-j@yandex.kz, dubovichenko@gmail.com

To estimate the contribution of (1)–(3) chains is possible *only* having the reliable data on *all* reaction rates. However, this information is not complete. So, for example, the reaction rate of the $^7$Li($^3$H,$n$)$^9$Be capture given in [8] have been obtained basing on the corresponding experimental cross sections. The neutron capture reaction rate on $^8$Li is calculated in [11]. The neutron capture reaction on $^{10}$B is considered in our work [12], and the neutron capture to the excited states (ESs) of $^{12}$B we examined in [13]. Besides, recently the radiative neutron capture reaction on $^{10}$Be [14] was studied. At a time the proton capture on $^7$Be [15] leading to $^8$B creation may be recognized as the beginning of the chain of synthesis of boron isotopes.

As it is seen from (1)–(3) the reaction $^{11}$B($n,\gamma$)$^{12}$B is the participant of all three chains. Therefore, in the present work the neutron capture on $^{11}$B to the ground state (GS) of $^{12}$B is considered, so as to include this reaction rate for estimation of processing of carbon isotopes according to chains (1)–(3). This is, in return, allows one to obtain the quantitative specifications of macroscopic parameters for the different BBN scenarios, considered in [16] in detail.

Thus, continuing to study the processes of radiative capture in the framework of the modified potential cluster model (MPCM) [17,18] we now consider the reaction of radiative capture $n^{11}$B → $^{12}$B at thermal and astrophysical energies. In the present research, we will use the new data on the spectra of $^{11}$B from [19], compared with our previous scientific papers [20], where the results of the earlier review [21] were used.

In addition, for the first time, resonances in the scattering phase shifts of the initial particles of the input channel are taken into account. In other words, the potentials of $n^{11}$B interactions for scattering processes now describe the main resonance states of the final nucleus $^{12}$B, which is considered in the $n^{11}$B channel at the positive energy. Despite the fact that there are no direct experimental measurements of total cross sections at energies greater than 70 keV [19,22], in the calculations of the $n^{11}$B → $\gamma^{12}$B reaction cross sections we took into account resonance states up to 5 MeV, which are now confirmed by new data on excitation energies and widths of the corresponding levels [19].

The $n^{11}$B → $\gamma^{12}$B reaction that we considered was studied earlier in [23], where the thermal cross section for neutron capture equal to 5.5 mb [24] was used to construct the $n^{11}$B interaction potentials. At the same time, in this paper we will use newer data from [25], where the value equals 9.09(10) mb is given.

Our calculations, in essence, are predictive and evaluative for substantiating and conducting new experiments of the neutron capture on $^{11}$B. Such results may also initiate a significant reassessment of the synthesis rate of the $^{12}$B isotope at temperatures above 0.2 $T_9$. This, in turn, will allow assessing how much the balance of elements synthesized in chains (1)–(3) can change, which is relevant for describing the evolution of astrophysical objects.

## 2. Calculation methods

In the last half century for the description of astrophysical reactions the microscopic model has become a successful direction – the resonant groups method (RGM) (see, for example, [26,27]), and also related models, for example, the generator coordinate method (GCM, see, in particular, [28]) or the algebraic version of the RGM (see, for example, [29] and references therein). However, the possibilities of simple potential two-cluster models (PCM) have not yet been fully investigated, especially if they use the concept of forbidden states (FSs) [30,31], and also directly take into account the behavior of the phase shifts of elastic scattering of interacting particles at low energies – such a variant of the model can be called modified PCM with FSs or MPCM [17,18].

In particular, in our previous works [11,14,15,17,18,32–34] (see also the references in the review [32]), the possibility of describing the basic characteristics of bound states (BSs) of some nuclei and astrophysical $S$-factors or total cross sections of radiative capture for more than 30 reactions of the type of radiative capture at thermal and astrophysical energies is shown. Calculations of these processes are made on the basis of the MPCM with the classification of states according to Young diagrams, which leads to forbidden, in some cases, states. The definite success of the MPCM can be explained by the fact that the potentials of intercluster interaction are built not only on the basis of known elastic scattering phase shifts, but with allowance for the classification of cluster states according to Young diagrams [30]. In addition, the potentials of bound or ground states with FSs in some partial waves are based on the description of the binding energy, radii and asymptotic constants (ACs) of the final nucleus in the cluster channel under consideration. Further, such potentials allow one to perform calculations of certain characteristics of the interaction of the considered particles in the processes of elastic scattering and capture reactions [17,18].

The total cross section of radiative capture $\sigma(NJ,J_f)$ for $EJ$ and $MJ$ transitions in the PCM are given in [35] or [17,18,36,37] and have the form

$$\sigma_c(NJ,J_f) = \frac{8\pi K e^2}{\hbar^2 q^3} \frac{\mu \cdot A_J^2(NJ,K)}{(2S_1+1)(2S_2+1)} \frac{J+1}{J[(2J+1)!!]^2} \sum_{L_i,J_i} P_J^2(NJ,J_f,J_i) I_J^2(J_f,J_i), \qquad (4)$$

where $\sigma$ is the total cross section of the radiative capture process, $\mu$ is the reduced mass of particles in the input channel, $q$ is the wave number of particles in the input channel, $S_1$ and $S_2$ are the spins of particles in the input channel, $K$ and $J$ is the wave number and the angular momentum of the γ-quantum in the output channel, respectively, $NJ$ are the $E$ or $M$ transitions of the $J$-th multipolarity from the initial $J_i$ state of two particles in the continuous spectrum to the final, bounded $J_f$ state of the nucleus in a two-cluster channel.

For electric orbital $EJ(L)$ transitions, the values of $P_J$, $A_J$ and $I_J$ have the form (see, for example, [35])

$$P_J^2(EJ,J_f,J_i) = \delta_{S_iS_f}\left[(2J+1)(2L_i+1)(2J_i+1)(2J_f+1)\right](L_i 0 J 0 | L_f 0)^2 \begin{Bmatrix} L_i & S & J_i \\ J_f & J & L_f \end{Bmatrix}^2,$$

$$A_J(EJ,K) = K^J \mu^J \left(\frac{Z_1}{m_1^J} + (-1)^J \frac{Z_2}{m_2^J}\right), \qquad I_J(J_f,J_i) = \langle \chi_f | r^J | \chi_i \rangle. \qquad (5)$$

Here $L_f$, $L_i$, $J_f$, $J_i$ are the orbital and total momenta of particles of the initial ($i$) and final ($f$) channels, $m_1$, $m_2$, $Z_1$, $Z_2$ are the masses and charges of the particles of the initial channel, $I_J$ is the integral over the wave functions of the initial $\chi_i$ and the final $\chi_f$ state of clusters, as functions of the relative motion of clusters with the intercluster distance $r$.

To consider the magnetic $MJ(S)$ transition, due to the spin part of the magnetic operator, we use the following expressions [17,18]

$$P_J^2(MJ, J_f, J_i) = \delta_{S_i S_f} \left[ S(S+1)(2S+1)(2J_i+1)(2L_i+1)(2L+1)(2J+1)(2J_f+1) \right] \cdot$$

$$\cdot (L_i 0 L 0 | L_f 0)^2 \begin{Bmatrix} L_i & L & L_f \\ S & 1 & S \\ J_i & J & J_f \end{Bmatrix}^2,$$

$$A_J(MJ, K) = \frac{\hbar K}{m_0 c} K^{J-1} \sqrt{J(2J+1)} \left[ \mu_1 \left(\frac{m_2}{m}\right)^J + (-1)^J \mu_2 \left(\frac{m_1}{m}\right)^J \right],$$

$$I_J(J_f, J_i) = \langle \chi_f | r^{J-1} | \chi_i \rangle, \text{ where } L = J - 1. \tag{6}$$

Here $m$ is the mass of the nucleus, $\mu_1$ and $\mu_2$ are the magnetic momenta of the clusters, the values of which are taken from the databases [38,39], namely $\mu_n = -1.91304272 \mu_0$ and $\mu(^{11}B) = 2.6887 \mu_0$.

Methods for calculating of other quantities in the MPCM framework, for example, root-mean-square mass and charge radii or binding energy, which are discussed further, are given, for example, in our works [17,18,37]. In the calculations we used the following values of particle masses $m_n = 1.00866491597$ amu [38], $m(^{11}B) = 11.0093052$ amu [39], and the constant $\hbar^2 / m_0$ was taken to be 41.4686 MeV·fm$^2$, where $m_0$ is amu.

## 3. Constructing potentials in the MPCM

Let us consider in more detail the procedure for constructing the intercluster partial potentials used here (in this case, the interaction potentials of the neutron and the target nucleus) for a given orbital angular momentum $L$ and other quantum numbers, determining the criteria and the algorithm of finding the parameters and indicating their errors and ambiguities [34].

First of all, parameters of potentials of bound states, for example, GS, should be determined, which, for a given number of bound states, are fixed quite uniquely by the binding energy, nuclear radius and AC in the channel under consideration. When constructing partial interaction potentials in MPCM, it is considered that they depend not only on the orbital momentum $L$, but also on the total spin $S$ and the total momentum $J$ of the considered system of nuclear particles. Here and in our previous works [17,18], we use Gaussian interaction potentials that depend on $JLS$ quantum numbers, as well as on Young diagrams.

In other words, for the different $JLS$ momenta we will have different values of the parameters of the partial potential. Since $E1$ or $M1$ transitions between different $^{(2S+1)}L_J$ states in the continuous and discrete spectra are usually considered, the potentials of these states will also be different [32,34]. Note once again that we use MPCM for our calculations and one of its modifications consists in the explicit dependence of the potentials used on Young diagrams {f} and taking into account, in some cases, the mixing of scattering states according to Young diagrams [17,18,30]. Taking into account the explicit dependence of the interaction potentials on Young diagrams makes it possible to use different potentials in scattering and discrete spectrum states if they depend on a different set of such diagrams, as was the case, for example, in the N$^2$H system [30,34].

The accuracy with which the parameters of the potential of the BSs are determined in this way is connected, first of all, with the accuracy of the AC. This potential does not contain any other ambiguities, since the classification of states according to Young diagrams makes it possible to unambiguously fix the number of bounded forbidden and allowed states in a given partial wave. Their number fully determines its depth, and the width of the potential depends entirely on the size of the AC. The principles for determining the number of FSs and allowed states (ASs) in a given partial wave based on Young diagrams for the $n^{11}$B system considered here are given below. They are described in more detail in [30,34].

It should be noted here that calculations of the charge radius in any model contain model errors. In any model, the magnitude of such a radius depends on the integral of the model wave functions (WFs), i.e. model errors of such functions are simply summarized. At the same time, AC values are determined by the asymptotic behavior of model WFs at one point and, apparently, contain a significantly smaller error. Therefore, the potentials of the BSs and GSs are constructed so that, first of all, they agree as much as possible with the values of AC, obtained on the basis of independent methods, which allow extracting AC from experimental data (see, for example, [40]).

The intercluster potential of the nonresonance process of scattering based on the phase shift of scattering for a given number of BSs allowed and forbidden in the considered partial wave is also can be constructed quite unambiguously. The accuracy of determining the parameters of such a potential depends, first of all, on the accuracy of extracting the scattering phase shifts from the experimental data and can reach 20–30%. Here, the potential does not contain ambiguities, since the classification of states according to Young diagrams makes it possible to unambiguously fix the number of BSs, which completely determines its depth, and the potential width at a given depth is determined by the shape of the scattering phase shifts.

When constructing a nonresonance scattering potential from the data on the spectra of the nucleus, it is difficult to estimate the accuracy of finding its parameters even for a given number of BSs, although it can be expected that it does not exceed the error in the previous case. Such a potential, as is usually assumed for the energy range up to 1–3 MeV, should lead to a scattering phase shift close to zero or give a smoothly falling form of the phase shift, since the resonance levels in the spectra of the nucleus are absent.

In the analysis of the resonance scattering, when in the considered partial wave at energies up to 1–3 MeV there is a relatively narrow resonance with a width of the order of 10–50 keV, for a given number of BSs, the potential is constructed distinctly. For a given number of BSs, its depth is completely fixed by the resonance energy of the level, and the width of the potential is determined by the width of this resonance. The error of its parameters usually does not exceed the error of determining the width of this level and is approximately 5–10%. Moreover, this applies to the construction of the partial potential from the resonance phase shifts of scattering and the determination of its parameters from the resonance in the spectra of the nucleus [34].

As a result, all these potentials do not contain ambiguities inherent in optical models [41], and as was shown [17,18] allow us to describe correctly the total cross sections of many processes of radiative capture. The potentials of the BSs should correctly describe the known values of the AC, which is associated with the asymptotic normalization coefficient ($A_{NC}$) usually extracted from the experiment as follows [40]

$$A_{NC}^2 = S_f \cdot C^2 \qquad (7)$$

where $S_f$ is the spectroscopic factor of the channel and $C$ is the dimensional AC, expressed in fm$^{-1/2}$ and determined from the relation

$$\chi_L(r) = C \cdot W_{-\eta L+1/2}(2k_0 r). \qquad (8)$$

$C$ is related to the dimensionless AC $C_w$ [42], used by us, as follows: $C = \sqrt{2k_0} C_w$, and the dimensionless constant $C_w$ can be defined by the expression [42]

$$\chi_L(r) = \sqrt{2k_0} \cdot C_w \cdot W_{-\eta L+1/2}(2k_0 r), \qquad (9)$$

where $\chi_L(r)$ is the numerical wave function of the bound state obtained from the solution of the radial Schrödinger equation and normalized to one, $W_{-\eta L+1/2}$ is the Whittaker function of the bound state, which determines the asymptotic behavior of the WF and is a solution of the same equation without nuclear potential, i.e. at large distances $r = R$, $k_0$ is the wave number due to the channel binding energy, $\eta$ is the Coulomb parameter, which is equal to zero in our case, and $L$ is the orbital momentum of this bound state.

## 4. State structure of the $n^{11}$B system

We do not have complete tables of products of Young diagrams for a system with more than eight particles [43], which we used earlier for similar calculations [17,18]. Therefore, the results obtained below should be considered only as a qualitative assessment of possible orbital symmetries in the GS of the $^{12}$B core for the $n^{11}$B channel. At the same time, based on this classification it was possible to adequately explain the available experimental data on the radiative capture of nucleons and light clusters on 1$p$-shell nuclei [17,18]. Therefore, here we will use the classification of cluster states according to orbital symmetries, which leads us to a certain number of FSs and ASs in partial intercluster potentials, and, therefore, to a certain number of nodes of the wave function of the relative motion of clusters - in this case, the neutron and $^{11}$B nucleus.

Further assume that for $^{11}$B (spin and isospin of $^{11}$B have the values $J^\pi, T = 3/2^-, 1/2$ [19]) we can accept the Young orbital diagram in the form {443}, therefore for the $n^{11}$B system we have {1} × {443} → {543} + {444} + {4431} [30,43]. The first of the diagrams obtained is compatible with the orbital momenta $L = 1,2,3,4$ and is forbidden, since there cannot be five nucleons in the $s$-shell [30,34], the second {444} diagram seems to be allowed and compatible with the orbital momentum $L = 0,2,4$, and the third one {4431}, also allowed, is compatible with $L = 1,2,3$ [31].

Thus, limited to only the lowest partial waves with the orbital momentum $L = 0,1,2,3$, we can say that for the $n^{11}$B system, only the AS for the {444} diagram is presented in the $S$ wave potential. In $P$ waves, there is a forbidden state at {543} and allowed at {4431}. In particular, in the $^3P_1$ wave with {4431} AS corresponds the GS of $^{12}$B with momenta $J^\pi, T = 1^+, 1$ at the binding energy of the $n^{11}$B system equals 3.370 MeV [19]. For $D$ waves, we have FS with the diagram {543} and AS with {4431} + {444}. For $F$ waves, we have FS with the diagram {543} and AS with {4431}. These ASs for the scattering potentials may be in the continuous spectrum and not be related. The methods used to determination the FSs or ASs allow us to determine their presence in a certain partial wave, but for the scattering states it is not possible to determine whether such a state will be bound. Therefore, for definiteness, we will further assume that the scattering states have only lower levels bound, whether they are allowed or forbidden.

If for $^{11}$B we accept the Young orbital diagram in the form {4421}, then for the $n^{11}$B system we have $\{1\} \times \{4421\} = \{5421\} + \{4431\} + \{4422\}$. The first diagram is forbidden - it corresponds to the orbital momenta $L = 1, 2, 3$, and the second one is allowed for the momenta $L = 1, 2, 3$, the third diagram is also allowed and has $L = 0, 2$. Thus considering only the waves with the orbital momentum $L = 0, 1, 2$, we can assume that for the $n^{11}$B system only the AS is present in the potential of the $^3S_1$ wave. In each $^3P$ wave there is a forbidden {5421} and allowed {4431} states. For $D$ waves, we have FS with the {5421} diagram and the AS for {4431} + {4422}, and for $F$ waves, we have the FS with the {5421} diagram and the AS for {4431}. ASs for scattering potentials can also be in the continuous spectrum, and not be bound. From this classification it is clear that in this case, the presence and number of bound FSs and ASs in these partial waves, as compared with the previous case, remains the same, although now they are compared with the other orbital Young diagrams.

Some $n^{11}$B-scattering states and BSs can be mixed along the channel spin with a value of 1 and 2. Therefore, both spin states $^3P_1$ and $^5P_1$ can contribute to the WF of the GS, and the GS should be considered as a $^{3+5}P_1$ mixture. This model does not allow to isolate states with $S = 1$ and 2 in the WF, so the GS function is $^{3+5}P_1$ level and is obtained by solving the Schrödinger equation with a given GS potential. A similar situation existed, for example, with neutron capture on $^{15}$N or $^7$Li, when the GS wave function was represented by a mixture of $^{3+5}P_2$ waves [17,18,44–46].

## 5. Possible transitions in the $n^{11}$B system

Since the GS of $^{12}$B here correlates the $^{3+5}P_1$ level, we can consider $E1$ transitions from the nonresonance $^3S_1$ and $^5S_2$ waves of the $n^{11}$B scattering at low energies to the different components of the WF of the GS of $^{12}$B in the $n^{11}$B channel

$$1. \begin{array}{c} ^3S_1 \xrightarrow{E1} {}^3P_1 \\ ^5S_2 \xrightarrow{E1} {}^5P_1 \end{array}. \tag{10}$$

The cross sections of this process can be written as a sum

$$\sigma(E1) = \sigma(^3S_1 \to {}^3P_1) + \sigma(^5S_2 \to {}^5P_1) \tag{11}$$

since the transitions originate from different partial scattering waves on the same GS of $^{12}$B, to which the spin-mixed $^{3+5}P_1$ WF is matched. Therefore, only coefficients $P_J$ will be different in the expression for the cross section (5) and (6).

Now consider excited states (ESs) of $^{12}$B that are bound in the $n^{11}$B channel

1. At an excitation energy of 0.95314(60) MeV or -2.41686(60) MeV [19] respect to the threshold of the $n^{11}$B channel, there is a first excited (1$^{st}$ ES) but bound in this channel state with the momentum is $J^\pi = 2^+$, which can be compared with the $^{3+5}P_2$ wave with the bound FS.

2. The second excited state (2$^{nd}$ ES) with an excitation energy of 1.67365(60) MeV

[21] relative to the GS or -1.69635(60) MeV relative to the threshold of the $n^{11}$B channel has $J^\pi = 2^-$ and it can be compared with the $^5S_2$ wave without a bound FS. In this case, a $^{3+5}D_2$ wave is also possible with the bound FS.

3. The third excited state (3$^{rd}$ ES) state with an excitation energy of 2.6208(12) MeV [19] or -0.7492(12) MeV relative to the threshold of the $n^{11}$B channel has $J^\pi = 1^-$ and can be compared with the triplet $^3S_1$ wave without the forbidden BS. In this case, a $^{3+5}D_1$ wave with a FS is also possible.

4. The fourth excited state (4$^{th}$ ES) with an excitation energy of 2.723(11) MeV or -0.647(11) MeV [19] with respect to the threshold of the $n^{11}$B channel has $J^\pi = 0^+$ and can be compared with a triplet $^3P_0$ wave with a bound FS.

Spectrum of the considered above ESs is shown in Fig. 1a

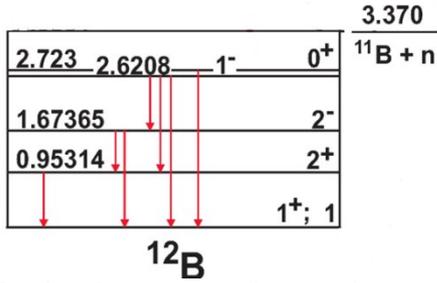

Fig.1a. Spectrum of excited states of $^{12}$B [19].

Furthermore, we construct the potentials of the last three ESs, which will be used to describe the scattering processes in these partial waves that do not contain resonances. However, there is a resonance in the $^{3+5}P_2$ scattering wave, and to build a $^{3+5}P_2$ potential, which has both the resonance and the first bound ES with the Gaussian form of potential parametrization used by us, is not possible. Therefore, in the future, when considering the capture to the first ES, different potentials for the discrete and continuous spectrum will be used. Here only $^{3+5}P_2$ potential for the resonance scattering wave will be obtained.

In addition to the excited states, there are several resonance states (RSs), i.e. states at positive energies relative to the threshold of the $n^{11}$B channel (italics indicate the levels that are present in the spectra, but are not considered by us for various reasons)

1. The first resonance state (1$^{st}$ RS) of $^{12}$B in the $n^{11}$B channel is at an excitation energy of 3.3891(16) MeV or $E_n$ = 20.8(5) keV, and has a width less than 1.4 keV in the c.m. and momentum $J^\pi = 3^-$ (see Table 12.10 in [19]) – it can be compared with the $^{3+5}D_3$ scattering wave with the forbidden bound state. We will consider this resonance, even only the *M*2 transition to the GS $^{3+5}P_1$ is possible for it.

2. The second resonance state (2$^{nd}$ RS) has energy $E_n$ = 430(10) keV – its width in the c.m. equal to 37(5) keV and the momentum $J^\pi = 2^+$ [19]. Therefore, it can be compared with the $^{3+5}P_2$ scattering wave with the bound FS and for it *M*1 transitions to the GS $^{3+5}P_1$ are possible.

3. The third resonance state (3$^{rd}$ RS) is at energy $E_n$ = 1027(11) keV with a small width equal to 9(4) keV in the c.m. and the momentum $J^\pi = 1^-$ [19] – it can be compared with the $^3S_1$ without the FS or $^{3+5}D_1$ scattering wave with the bound FS and for it *E*1 transitions to the GS $^{3+5}P_1$ are possible.

*4. The fourth resonance state (4$^{th}$ RS) is at energy $E_n$ = 1.19 MeV with an unknown, but large width and momentum 2$^-$ (see table 12.10 from [19]) – it can be compared to $^5S_2$ or $^{3+5}D_2$ scattering wave and for it E1 transitions to the GS $^{3+5}P_1$ are possible. However, we will not consider these transitions due to the unknown width.*

*5. The fifth resonance (5$^{th}$ RS) has energy $E_n$ = 1.280(20) MeV with a width of 130(20) keV in the c.m. and the momentum 4$^-$ [19]. We will not consider it, because here the E1 or M1 transitions to the GS are impossible.*

6. The sixth resonance (6$^{th}$ RS) with $E_n$ = 1.780(20) MeV with a width 60(20) keV in the c.m. and momentum $J^\pi = 1^+$ can be compared with the $^{+5}P_1$ scattering wave with the bound FS (see Table 12.10 [19]) and $M$1 transitions to the GS $^{3+5}P_1$ are possible for it. This state can be attributed also to the $^5F_1$ wave with the FS, but then only the $E$2 transition to the GS is possible. For it, we will build the resonance potential and determine the contribution of this process to the total capture cross sections.

7. The seventh resonance (7$^{th}$ RS) at $E_n$ = 2.450(20) MeV with a width 110(40) keV in the c.m. and the momentum $J^\pi = 3^+$ can be compared with either the $^5P_3$ or $^{3+5}F_3$ scattering waves (see Table 12.10 from [19]). We will consider it, and determine the contribution of this an $E$2 process of the total capture to the GS cross sections.

8. *The eighth resonance (8$^{th}$ RS) at $E_n$ = 2.580(20) MeV with a width 55(20) keV in the c.m. and the momentum $J^\pi = 3^-$ can be compared with the $^{3+5}D_3$ scattering wave (see Table 12.10 from [19]) and we will not consider it, since only the M2 process is possible here.*

9. *The ninth resonance (9$^{th}$ RS) with $E_n$ = 2.9 MeV with unknown width and momentum $J^\pi = 1^-$ can be compared with the $^3S_1$ scattering wave without FS (see table 12.10 from [19]) and can lead to E1 transitions to the GS. However, we will not consider it because of the unknown width.*

10. The tenth resonance (10$^{th}$ RS) at $E_n$ = 3.5 MeV with a width of 140 keV and a momentum $J^\pi = 1^+$ can be compared with the $^{3+5}P_1$ scattering wave with the bound FS (see Table 12.10 from [19]) and can lead to $M$1 transitions on the GS. This state can be also attributed to the $^5F_1$ wave with the FS, but then only the $E$2 transition to the GS is possible. For it, we construct a resonance potential and determine the contribution of this process to the total capture cross sections.

11. *Next comes the resonance (11$^{th}$ RS) at a neutron energy of 4.03 MeV with an unknown width and momentum 1$^-$. It can be compared with the $^3S_1$ wave without the FS, and E1 transitions to the GS $^{3+5}P_1$ are possible for it. We will not consider it because of the unknown width.*

12. *The next resonance (12$^{th}$ RS) is at energy of 4.55 MeV with a width of less than 14 keV in the c.m. and an unknown momentum. However, we will not consider it because of the unknown momentum.*

13. Next is the resonance state (13$^{th}$ RS) with a neutron energy of 4.70 MeV, a width of 45 keV in the c.m. and a momentum 2$^-$. It can be compared with the $^5S_2$ wave without the FS, and $E$1 transitions to the GS $^{3+5}P_1$ are possible for it. In this case, a $^{3+5}D_2$ wave with the FS is also possible, which allows for an $E$1 transition to the GS.

14. Resonance state (14$^{th}$ RS) at 4.80 MeV energy with a width of 90 keV in the c.m. has a momentum 1$^-$. It can be compared with the $^3S_1$ wave without the FS, and $E$1 transitions to the GS $^{3+5}P_1$ are possible for it. In this case, the $^{3+5}D_1$ wave with the FS is also possible, which allows for the $E$1 transition to the GS.

15. Higher states have not yet been studied in a detail [19], and we will not consider them. As a result, the influence of 7 resonances at energies up to $E_n$ = 5 MeV can be considered – these are resonances number 2, 3, 6, 7, 10, 13 and 14. The rest have no either known width or momentum, so it is impossible for them to build unambiguous potentials, as is done for the remaining resonances. Some resonances, for example, the 1$^{st}$ RS, do not lead to the $E$1 or $M$1 transitions and will not be considered. In addition, it will be shown that when considering the $E$2 transitions for resonances 6, 7 and 10, if we accept $F$ waves for them, the cross sections obtained are several orders of magnitude smaller than the cross sections for the $E$1 and $M$1 transitions. Spectrum of the described resonance states is shown in Fig. 1b [19].

| $E_n$ (MeV ± keV) | $\Gamma_{cm}$ (keV) | $^{12}B^*$ (MeV) | $l$ | $J^\pi$ |
|---|---|---|---|---|
| 0.0208 ± 0.5 | ≪ 1.4 | 3.3889 | 2 | 3⁻ |
| 0.43 ± 10 | 37 ± 5 | 3.764 | 1 | 2⁺ |
| 1.027 ± 11 | 9 ± 4 | 4.311 | 0 | 1⁻ |
| 1.19 | broad | 4.46 | 0, 2 | 2⁻ |
| 1.28 ± 20 | 130 ± 20 | 4.54 | 2 | 4⁻ |
| 1.78 ± 20 | 60 ± 20 | 5.00 | 1 | 1⁺ |
| 2.45 ± 20 | 110 ± 40 | 5.62 | 1 | 3⁺ |
| 2.58 ± 20 | 55 ± 20 | 5.73 | 2 | 3⁻ |
| 2.9 | broad | 6.0 | 0, 2 | 1⁻ |
| 3.5 | 140 | 6.6 | 1 | 1⁺ |
| 4.03 | broad | 7.06 | 0, 2 | 1⁻ |
| 4.55 | ≤ 14 | 7.54 | > 3 | |
| 4.70 | 45 | 7.68 | 0, 2 | 2⁻ |
| 4.80 | 90 | 7.77 | 0, 2 | 1⁻ |

Fig. 1b. Spectrum of the resonance states of $^{12}$B from Table 12.10 [19] at energies down to 5 MeV.

As it was shown above, resonance states 3 and 14 coincide with the 3$^{rd}$ ES by their momentum and the 13$^{th}$ RS with the second one. However, it is impossible to construct *S*-potentials that would have a bound AS coinciding with one of the ESs and having resonance with the observed excitation energy. Therefore, we construct these resonance potentials so that they correspond to *D* waves with the FS and have a resonance at the required energy with the required width. Based on the given above information about the ESs and RSs, it is possible to consider other *E*1 transitions from nonresonance in the considered region of the energy *S* and *D* scattering waves, as well as *M*1 processes from the resonance $^{3+5}P_2$ wave and non-resonance $P_0$ wave. As already mentioned, for the *S* waves and $P_0$ scattering wave we will use the potentials of the 2$^{nd}$, 3$^{rd}$ and 4$^{th}$ ESs, and for the 2$^{nd}$ RS in the $P_2$ wave potential, which has a resonance and a bound FS, and does not coincide with the 1$^{st}$ ES. As a result of the analysis of these RSs, the transitions that are listed in Table 1 will be further considered.

Cross sections of some *E*1 transitions, for example, 3, 5 and 6 from Table 1 can be written as

$$\sigma(E1) = \{\sigma(^3D_1 \to ^3P_1) + \sigma(^5D_1 \to ^5P_1)\}/2 + \{\sigma(^3D_2 \to ^3P_1) + \sigma(^5D_2 \to ^5P_1)\}/2 + \\ + \sigma(^5D_0 \to ^5P_1) \quad . \quad (12)$$

Here, the cross section averaging is performed for the transition from the mixed $D_1$ and $D_2$ scattering states to the mixed GS. Based on the observables (for resonances this is the state energy and its width), only the potential of $^{3+5}D_1$ or $^{3+5}D_2$ waves can be built, as well as the $^{3+5}P_1$ potential for the GS. Therefore, transitions 3 or 5 differ only in the spin coefficients in the expression for cross section (5), and the matrix elements are calculated between the same WFs mixed along the spin of the channel — this question is described in more detail in [11,17,18].

The cross section for the *M*1 processes Nos. 7 and 8 is written in the form

$$\sigma(M1) = \sigma(^3P_0 \to ^3P_1) + \{\sigma(^3P_2 \to ^3P_1) + \sigma(^5P_2 \to ^5P_1)\}/2, \quad (13)$$

because the mixed $^{3+5}P_2$ wave is used in the initial channel.

Table 1. A list of possible transitions from the initial $\{^{(2S+1)}L_J\}_i$ state to the different components of the WF $\{^{(2S+1)}L_J\}_f$ of the neutron capture on $^{11}$B for the GS of $^{12}$B and the parameters of Gaussian potentials, which are defined below, for the initial states. The value of $P^2$ determines the coefficient in the cross sections (5) and (6). The number of resonances from the list of RSs given above is given.

| No. | $\{^{(2S+1)}L_J\}_i$ for the initial $n^{11}$B channel | Transition type | $\{^{(2S+1)}L_J\}_f$ for the final $n^{11}$B channel | $P^2$ (to define the cross sections) | $V_0$, MeV | $\alpha$, fm$^{-2}$ |
|---|---|---|---|---|---|---|
| 1. | $^3S_1$ nonresonance scattering wave. | E1 | $^3P_1$ | 3 | 5.61427 | 0.04 |
| 2. | $^5S_2$ nonresonance scattering wave. | E1 | $^5P_1$ | 3 | 13.56295 (6.70125) | 0.1 (0.03) |
| 3. | $^3D_1$ resonance at 1.027 MeV – No. 3 | E1 | $^3P_1$ | 3/2 | 1611.95103 | 1.25 |
|  | $^5D_1$ resonance at 1.027 MeV – No. 3 | E1 | $^5P_1$ | 27/10 |  |  |
| 4. | $^3D_1$ resonance at 4.8 MeV – No. 14 | E1 | $^3P_1$ | 3/2 | 4502.245 | 3.5 |
|  | $^5D_1$ resonance at 4.8 MeV – No. 14 | E1 | $^5P_1$ | 27/10 |  |  |
| 5. | $^3D_2$ resonance at 4.7 MeV – No. 13 | E1 | $^3P_1$ | 9/2 | 6444.382 | 5.0 |
|  | $^5D_2$ resonance at 4.7 MeV – No. 13 | E1 | $^5P_1$ | 21/10 |  |  |
| 6. | $^5D_0$ nonresonance scattering wave. | E1 | $^5P_1$ | 6/5 | 800 | 1.0 |
| 7. | $^3P_2$ resonance at 430 keV – No. 2 | M1 | $^3P_1$ | 5/2 | 11806.017 | 15.0 |
|  | $^5P_2$ resonance at 430 keV – No. 2 | M1 | $^5P_1$ | 9/2 |  |  |
| 8. | $^3P_0$ nonresonance scattering wave. | M1 | $^3P_1$ | 2 | 147.1709 | 0.18 |
| 9. | $^3P_1$ nonresonance scattering wave. | M1 | $^3P_1$ | 3/2 | 194.68751 (44.70853) | 0.22 (0.12) |
|  | $^5P_1$ nonresonance scattering wave. | M1 | $^5P_1$ | 27/2 |  |  |
| 10. | $^3F_3$ resonance at 2.45 MeV – No. 7 | E2 | $^3P_1$ | 6 | 5981.83 | 0.32 |
|  | $^5F_3$ resonance at 2.45 MeV – No. 7 | E2 | $^5P_1$ | 54/25 |  |  |
| 11. | $^5F_1$ resonance at 1.78 MeV – No. 6 | E2 | $^5P_1$ | 81/25 | 5145.881 | 0.275 |
| 12. | $^5F_1$ resonance at 3.5 MeV – No. 10 | E2 | $^5P_1$ | 81/25 | 8787.42 | 0.47 |
| 13. | $^{3+5}D_3$ resonance at 20.8 keV – No. 1 | M2 | $^3P_1$ | 21/5 | 16.0577867 (64.59402235) | 0.0125 (0.05) |
|  |  |  | $^5P_1$ | 14/5 |  |  |

It should be noted here that $M1$ transitions from the $^{3+5}P_1$ scattering wave to the $^{3+5}P_1$ GS in one potential result in zero matrix elements (ME). In other words, if we use the $^{3+5}P_1$ GS potential for the $^{3+5}P_1$ scattering wave, the phase shifts of which are shown in Fig. 1a, then the ME of such transitions is zero. At once, we say that in numerical terms, the total cross section for such a transition with the same potential does not exceed $10^{-7}$ μb, and the very minimum of the cross sections in our further calculations is of the order of 5 μb. Further, we use precisely these versions of potentials and transition No. 9 makes no real contribution to the total cross sections. Moreover, if we use for a $^{3+5}P_1$ scattering wave a potential, leading, for example, to zero scattering phase shifts, this transition leads to a value of about 1.5 μb at 7 MeV, which gradually increases at zero energy from 0.2 μb, which is about 1.5% of total cross section at maximum energy. In this case, the contribution of this transition is small and it does not give a real effect on the total cross sections.

## 6. $n^{11}B$ interaction potentials

For all partial potentials, i.e. interactions for each orbital moment $L$ at given $JS$, a Gaussian type was used

$$V(^{2S+1}L_J, r) = -V_0(^{2S+1}L_J)\exp\{-\alpha[^{2S+1}L_J]r^2\], \tag{14}$$

where the depth $V_0$ and the width $\alpha$ of the potential depend on the momenta $^{2S+1}L_J$ of each partial wave. In some cases, this potential may also depend explicitly on the Young diagrams $\{f\}$ and be different in the discrete and continuous spectrum, since in such states these diagrams are different [30].

Furthermore, to build the potential of the GS, we note that in [4,47–49] for the $A_{NC}$ of the form (7) the values of 1.13 to 1.35 fm$^{-1}$ were obtained, and in [4,19,23,50], the spectroscopic factor of the GS is given in the range of 0.69 to 1.30. As a result, for the range of dimensional AC we get 0.87–1.96 fm$^{-1}$ or 0.93–1.40 fm$^{-1/2}$, which for the dimensionless AC $C_w$ of the form (9) with $\sqrt{2k_0} = 0.880$ gives an interval of 1.06–1.59 with an average value of 1.32(27).

We now build the potential of the GS with the FS, which corresponds to the AC from this range – its parameters are given under No. 9 in Table 1. This potential gives a negative AC -1.64(1), because it contains the FS, that is stable in the range of 5–17 fm, a charge radius of 2.43 fm and a mass radius of 2.53 fm at a binding energy of -3.37000 MeV with an accuracy of $10^{-5}$ MeV [17,18,34], which completely coincides with the experimental value [19]. Hereinafter, the error of the AC is determined by its averaging over a specified distance interval, and the scattering phase shift of this potential is shown in Fig. 1a by the blue solid curve. One can use, for example, the matter radius of 2.41(3) fm obtained in [51] for comparison.

For the third excited $^3S_1$ level with an excitation energy of 2.62 MeV or -0.749 MeV relative to the threshold of the $n^{11}B$ channel [19], the size of the $A_{NC}$ (with $S_f = 1$) is given in [4,47,48] $A_{NC} = C = 0.94(8)$ fm$^{-1/2}$. The coefficient $\sqrt{2k_0}$ for this level is 0.605, so for the $C_w$ from (8) we get the interval 1.42–1.69 with an average value of 1.55(14). In a new paper [52], $S_f = 0.63$ and $A_{NC} = -1.05(5)$ fm$^{-1/2}$ are given, then we have $C_w = 2.2(1)$. If we further consider that the AS for the {444} diagram in the $^3S_1$ wave can be bound and correspond to the third excited state of $^{12}B$ at an energy of -0.7492 MeV relative to the channel threshold, then for the parameters of the $^3S_1$ potential without FS, one can get the values given under No. 1 in Table 1.

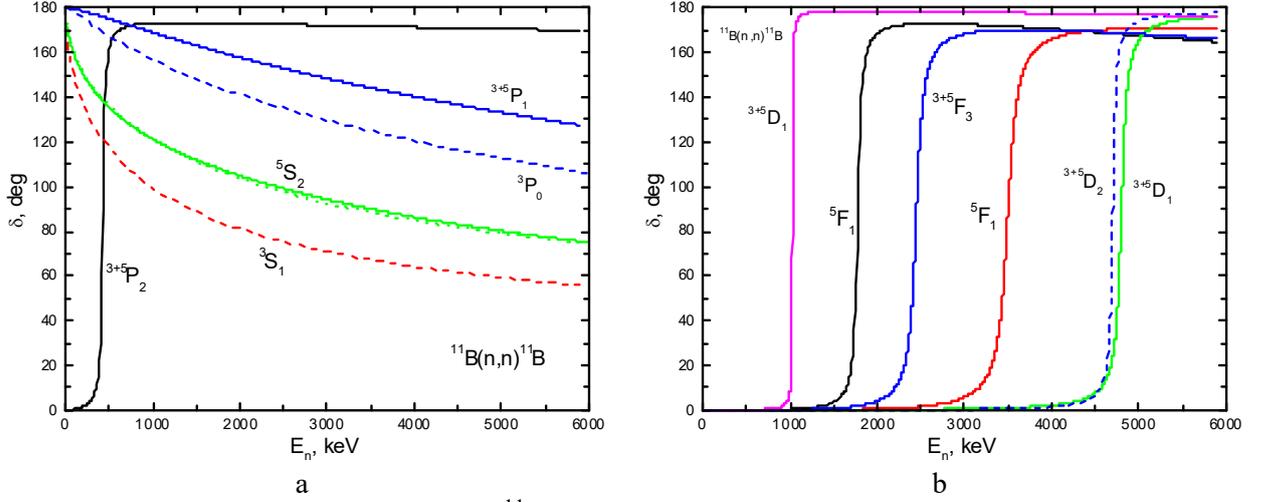

Fig. 2a,b. Phase shifts of the elastic $n^{11}$B scattering at low energies obtained using potentials with parameters from the text.

Using this potential, the binding energy of -0.7492 MeV at $\varepsilon = 10^{-4}$ [34] was found, which fully coincides with the experimental value [19], the charge radius is 2.47 fm, the mass radius equals 2.94 fm and the dimensionless AC equal to +1.9(1) over the interval 10–30 fm. The form of the $^3S_1$ scattering phase shift with this potential is shown in Fig. 2a by the red dashed curve. This potential will be used further to describe the $^3S_1$ scattering processes and to calculate the first transition under No. 1 of Table 1. The value of its AC is approximately in the middle of the estimated intervals obtained from the data of [4,47,48,52].

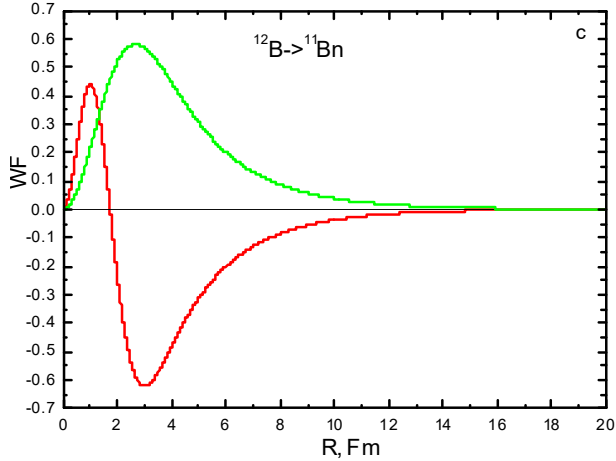

Fig. 2c. WF of the GS of $^{12}$B in the $n^{11}$B channel.

For the 2$^{nd}$ ES $^5S_2$, the value 1.34(12) fm$^{-1/2}$ (at $S_f = 1$) was obtained in [4,47,48], which, at a binding energy of -1.6964 MeV and $\sqrt{2k_0} = 0.742$, gives $C_W$ an interval of 1.64–1.97 with an average value of 1.81(17). In a newer paper [52], the values $S_f = 0.33$ and $A_{NC} = -1.28(6)$ fm$^{-1/2}$ are given, then we have $C_W = 3.0(1)$. In this case, and by analogy with the $^3S_1$ wave, for the parameters of this potential you can get the values of No. 2 from Table 1. This potential leads to a binding energy relative to the channel threshold of -1.6964 MeV, AC equal to +1.9(1) over the interval 6–26 fm, charge radius equals 2.53 fm and mass radius is 2.60 fm. The form of the $^5S_2$ scattering phase shift with this potential is shown in Fig. 2a by the green dotted curve. This potential also will be used further to describe the $^5S_2$ scattering processes and calculate transition No. 2 from Table 1.

This potential has the AC value noticeably less than the results of work [52]; therefore, we consider the second version of this potential No. 2, given in Table 1 in brackets. It leads to the same binding energy, gives the AC equal to +3.1(1), the mass radius is 2.78 fm and the charge radius is 2.45 fm, and its phase shift is shown in Fig. 2a by the green solid curve, which is practically the same as the green dotted curve for the previous potential. However, as will be seen later, its impact on the final results is very small.

It should be noted here that in paper [52] for both $S$ waves the negative sign of AC is given. For the 3$^{rd}$ ES it is -1.05(5) fm$^{-1/2}$ and for the 2$^{nd}$ ES it is -1.28(6) fm$^{-1/2}$. The negative sign of AC means the presence of a node in the wave function (WF) of the relative motion of the clusters and the work [52] gives the form of these WFs. However, according to our classification, there are no bound forbidden states in the $S$ waves. According to the classification of states using Young diagrams given below, the bound FS is present only in the $P$ waves of the discrete spectrum. Therefore, for the GS with potential No. 9 from Table 1, the AC has a negative sign of -1.64(1). The WF of this potential is shown in Fig. 2a by the red solid curve. The green solid curve shows the WF of the GS of $^{12}$B in the same channel for the potential without the bound FS with the parameters shown in Table 1 No. 9 in brackets. This potential leads to a binding energy of -3.37000, the same, but positive, AC +1.64, mass radius of 2.53 fm, and charge radius of 2.42 fm. The results of the calculation of the total cross sections with this potential and their comparison with the previous potential of the GS will be given below.

For the potential of the resonance $^{3+5}P_2$ wave with one bound FS, parameters No. 7 were obtained. This potential leads to a resonance energy $E_n$ = 430(1) keV with a width of 37(1) keV (cm), which completely coincide with the experimental data [19] – for this energy, the scattering phase shift turned out to be 90(1). To calculate the level width with the scattering phase shift δ the expression Γ = 2($d\delta/dE$)$^{-1}$ was used, and the form of the resonance mixed $^{3+5}P_2$ scattering phase shift is shown in Fig. 2a by the black solid curve. Recall that the exact mass values of the particles and the $\hbar^2/m_0$ constant value that is given above were used to calculate the scattering phase shifts. These values strongly influence the position of the resonance under consideration or the BS binding energy in the $n^{11}$B channel.

For the potentials of a $^{3+5}P_1$ scattering wave, we will use the potential of the GS No. 9, and for a nonresonance $^3P_0$ scattering wave with the bound FS, we will use the potential that is matched by the 4$^{th}$ ES, by analogy with the previous ESs No. 1 and No. 2 in Table 1. Based on the results of [52], where $A_{NC}$ = 0.15(1) and $S_f$ = 0.113 are given, it is possible to obtain with $\sqrt{2k_0}$ = 0.583 the dimensionless AC equals 0.77(2) and potential parameters, which are given under No. 8 in Table 1. This potential gives a binding energy of -0.6470 MeV, a negative AC $C_w$ = -0.75, because it contains FS, the mass radius is 2.75 fm and the charge radius is 2.45 fm. Its scattering phase shift is shown in Fig. 2a by the blue dashed curve.

To consider the $E$1 transitions from the nonresonance $^5D_0$ scattering wave with the bound FS on the GS, one can use the parameters No. 6 in Table 1, which lead to the scattering phase shifts less than 0.1° at energies up to 6 MeV in a laboratory system. Here we must remember that if the potential contains $N + M$ forbidden and allowed states, it obeys the Levinson generalized theorem and its phase shift at zero energy begins with π·($N + M$) [30]. However, in Fig. 2a, the $P_1$ phase shift, having the bound FS and the bound AS, is from 180°, not from 360°, and the $P_2$ phase shift, containing one bound FS, starting from 0 degrees, but not from 180° for a more familiar presentation of the results and placement of all phase shifts in one figure. Only the $P_0$-phase shift with one bound FS is shown correctly. $S$-phase shifts, the potentials of which do not have bound FSs or ASs, are shown from 180°, and not from 0, for the same reasons. The same comments also apply to Fig. 3a.

For $^{3+5}D_2$, $^{3+5}D_1$ resonances at 4.7 and 4.8 MeV, respectively, potentials with the FSs No.5 and No.4 were obtained. The first of them gives a width of 49(1) keV, and the second one is of 86(1) keV, which agrees well with the data [19], and the energies exactly coincide with those given above. The phase shifts of these potentials are shown in Fig. 2b by the blue dashed and green solid curves – at resonance they have a value of 90(1)°.

Resonance at 1.027 MeV was also able to be reproduced only under the assumption of the $^{3+5}D_1$ wave and parameters No. 3 with the FS, which lead to a width of 8.8 keV with exactly coincident energy and scattering phase shift, which is shown in Fig. 2b by the violet solid curve, which in resonance is 90(1)°.

The resonances at 1.78 and 3.5 MeV could be reproduced only under the assumption of $F$ waves. In particular, for $^5F_1$ waves, the parameters with the FS No. 11 and No. 12 are obtained. The first of these gives a width of 56(1) keV, and the second of 149(1) keV with exactly the same energy and scattering phase shift shown in Fig. 2b by the black and red solid curves, which in resonance have values of 90(1)°.

The resonance at 2.45 MeV was also able to be reproduced only under the assumption of $F$ waves. For the $^{3+5}F_3$ wave, we obtained the parameters with the FS No.10, which lead to a width of 110(1) keV with exactly the same energy and the scattering phase shifts shown in Fig. 2b by the blue solid curve, which has a resonance value of 90(1)°.

Furthermore, the potential for the first resonance state was obtained – it is listed in Table 1 under No. 13. Such potential leads to the resonance energy of 20.80(1) keV at width of 0.5(1) keV. It should be noted that there is no exactly measured width for this resonance and in [19] the width value lower 1.4 keV is given. Therefore, there is very large parameter values range, which allow one to obtain the width value lower 1.4 keV, but lead to not so very small difference in results. Therefore, other option of potential parameters for this potential is given for comparison, which is shown in Table 1 under No.13 in curly brackets. It leads to the resonance at 20.80(1) keV with the width of 0.07(3) keV, which is also smaller 1.4 keV [19].

## 7. The $^{11}$B$(n,\gamma)^{12}$B capture total cross sections and reaction rate

The calculation results of the total cross sections for the first transition from Table 1 with the scattering potential to the GS with potential No. 9 from Table 1 are shown in Fig. 3a by the green dotted curve. These results describe well the experimental data of [25] at the energy of 25.3 meV (black square) and in the region of 23 ± 61 keV (black circles) [53] for the capture to the GS with the potential No. 9 from Table 1. The cross sections were calculated to neutron energy up to 7 MeV, but do not take into account resonances above 5 MeV. Calculation results of the cross section for the transition No. 8 from the nonresonance $^3P_0$ scattering wave to the GS. The potential of the 4$^{th}$ ES, which is coordinated with AC from [52] is used for $^3P_0$ scattering wave. The value of this cross section in the maximum is at the level 0.15 μb and is not shown in Fig. 3a. If we use for it the parameters leading to zero scattering phase shifts, the cross section turns out to be much smaller. Theoretical estimation of cross section errors does not exceed 5–10% at worst. Experimental cross section errors are shown in this paper subject to these data.

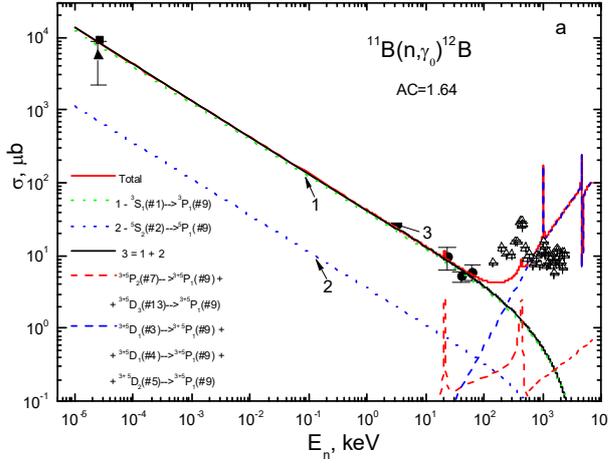
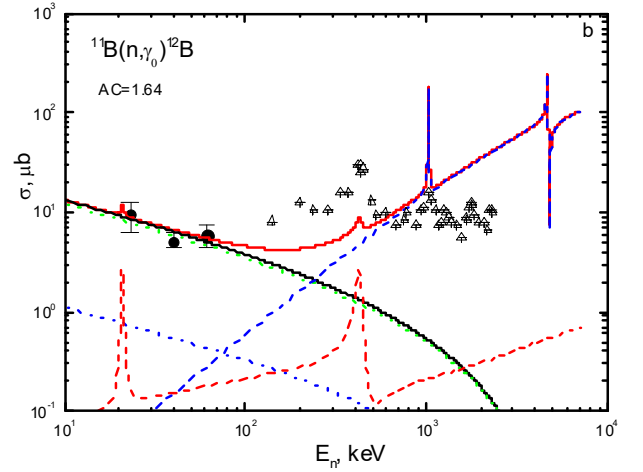

Fig.3a. Total cross sections of the neutron radiative capture reaction on $^{11}$B for transition to the GS in the energy range of $10^{-5}$ keV to 7 MeV for the GS potential with AC equals 1.64. Experimental data: black triangles (▲) – data from [24] at thermal energy, black square (■) – data from [25] at thermal energy, black points (●) – total cross sections to the GS [53], open triangles (Δ) – total cross sections [54]. Curves – calculation for different transitions (Table 1) with the potentials given in the text. The numbers of potentials from Table 1 are given in the legend of Fig.3a.

Fig.3b. Total cross sections of the neutron radiative capture reaction on $^{11}$B for transition to the GS in the energy range of 10 keV to 7 MeV.

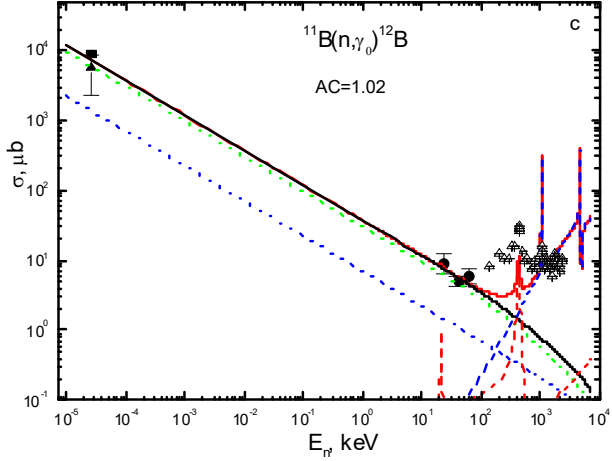

Fig.3c. Total cross sections of the neutron radiative capture reaction on $^{11}$B for transition to the GS in the energy range of $10^{-5}$ keV to 7 MeV for the GS potential with AC equals 1.02. Notations as in Fig. 3a.

If we assume that the GS is the $^{3+5}P_1$ mix, then we can consider the second $E1$ transition under No. 2 from Table 1, still using the GS potential No. 9 from Table 1, which is the potential of the mixed $^{3+5}P_1$ GS of $^{12}$B in the $n^{11}$B channel. The results of the calculation of the total cross section are shown in Fig. 3a by the blue dotted curve, which leads to the cross sections more than an order of magnitude smaller than the green dotted curve for transition No. 1 from Table 1. The total cross section for both processes No. 1 is shown in Fig. 3a by the black solid curve. If parameters from brackets under No. 2 in Table 1 for the $^5S_2$ wave potential are used, which are consistent with the AC data from [52], the result for the cross sections will differ from the blue dotted curve upwards by about 3%. We will not consider this result here, considering these changes as not significant.

The red dashed curve in Fig. 3a shows the result of calculating the cross sections of the $M$1 capture from the resonance $^{3+5}P_2$ wave No. 7 from Table 1 and the $M$2 capture from the resonance $^{3+5}D_3$ wave No. 13 from Table 1. The blue dashed curve shows the $E$1 cross sections with regard to $^{3+5}D_1$ resonances No. 3 at 1027 keV, $^{3+5}D_1$ No. 4, and $^{3+5}D_2$ No. 5 with resonances at 4.7 and 4.8 MeV. The cross sections of resonances No. 10, No. 11, and No. 12 have a very small size and are not visible in Fig. 3a. The red solid curve shows the summarized total cross sections for all processes listed above. More thoroughly all resonances are shown in Fig. 3b.

As it was mentioned, the resonance at 20.8 keV with potential No. 13 from Table 1, which allows one to consider $M$2 transition to the GS, is studied – results are shown in Fig. 3a by the red dashed curve. It leads at energy of 20.8 keV to the peak of total cross sections with the value of about 2.6 µb at summed cross sections 11.8 µb. This peak is seen in Fig. 3a near first point of [53] at 23 keV and is in the limit of error of experimental data. If to use the potential, which parameters are given in curly brackets under No. 13, then the resonance value rises up to 58 µb at the resonance in total cross section of 67 µb. There is no such sharp rise in the experiment; therefore the first option of potential No. 13 is more preferable. The red solid line shows summed total cross sections for all above mentioned processes. More thoroughly all resonances are shown in Fig. 3b.

As can be seen from Figs. 3a,b taking into account only transitions from the $S$-scattering waves allows one to correctly describe the available experimental data in the range of 25 meV to 60 keV. The total cross sections in this energy region are well described by an almost straight line, which at 25 meV agrees with the new data from [25]. The structure of resonances obtained above 400 keV can be considered as a prediction of the results for the total cross sections and a stimulus for new experimental studies of this reaction.

If we use the potential parameters for the GS of $^{12}$B, shown in brackets under No. 9 in Table 1, the total cross sections, while retaining their shape, turn out to be an order of magnitude higher. They are no longer able to correctly describe either the data at 25 meV [25] or the measurement results at 20–60 keV from [53]. All resonances are preserved, but they have an order of magnitude or more than the previous results. This value of the total cross sections follows from the shape of the WF of the GS – since the ME is an integral of the WF of scattering and BS, then this integral will be noticeably larger when using the WF shown in Fig. 2c by the green solid curve.

Calculation results for the GS potential, which leads to the AC of 1.02 and has parameters $V_0 = 419.108548$ MeV and $\alpha = 0.5$ fm are shown in Fig. 3c. Such potential leads to the mass radius of 2.45 fm and the charge radius of 2.42 fm. As seen from this figure the total cross section at 25.3 meV has slightly smaller value comparably with first GS potential No. 9 from Table 1 and, as before, describes experimental results of work [53] shown in Fig. 3c by dots. This two GS potentials lead to the AC at the limit of possible range of their values, as it was shown above.

Since, as can be seen from Fig. 3a, at energies of 10 meV to 10 keV, the calculated cross section is practically a straight line, it can be approximated by a simple function of the form

$$\sigma_{ap}(\mu b) = \frac{A}{\sqrt{E(\text{keV})}}. \tag{15}$$

The value of the constant $A = 40.82$ μb·keV$^{1/2}$ was determined by one point in the calculated cross sections (red solid curve in Fig. 3a) with a minimum energy of 10 meV. The value of the cross section is equal to 8.4 mb at thermal energy of 25.3 meV. The module

$$M(E) = |[\sigma_{ap}(E) - \sigma_{theor}(E)]/\sigma_{theor}(E)| \tag{16}$$

of the relative deviation of the calculated theoretical cross section ($\sigma_{theor}$) and the approximation ($\sigma_{ap}$) of this cross section by the given above function in the region up to 10 keV does not exceed 0.5%. The constant value $A$ is equal to 36.695 μb·keV$^{1/2}$ for Fig. 3c at the same error at 10 keV. The total cross value at thermal energy now equals 7.3 mb.

Furthermore, the reaction rate of the neutron capture on $^{11}$B was calculated, and in units of cm$^3$ mol$^{-1}$s$^{-1}$ it can be represented as [35]

$$N_A\langle\sigma v\rangle = 3.7313 \cdot 10^4 \mu^{-1/2} T_9^{-3/2} \int_0^\infty \sigma(E) E \exp(-11.605 E/T_9) dE, \tag{17}$$

where $E$ is given in MeV, the total cross section $\sigma(E)$ is measured in μb, $\mu$ is the reduced mass given in amu, and $T_9$ temperature is given in 10$^9$ K [35]. To obtain this integral, the calculated total cross section for 7000 points in the energy range of 10$^{-5}$ to 7·10$^3$ keV was used.

For the reaction rate, these results are shown in Fig. 4 – the notation for the curves is similar to Fig. 3a (*green and blue dotted curves, black and red solid curves*). As can be seen from Figs. 3a and 4, taking into account resonances at energy of about 5 MeV strongly influences the total cross sections in this area, leading to their large rise, and greatly affects the reaction rate, increasing its value by about 20 times at 10 $T_9$. Moreover, this increase is already starting to affect at temperatures above 0.2–0.3 $T_9$.

For comparison, the blue solid curve shows the results from [23]. The calculation of this rate was performed taking into account the resonances and, presumably, the capture for all ESs, and the contribution of the direct capture process [23] is shown in Fig. 4 by the blue dashed-dotted curve [23], which is generally consistent with our results. A small, approximately one and a half times, excess of our reaction rate over the results of [23] can be explained by the fact that in this work the cross section was normalized to the data of [24], i.e. to the magnitude of the thermal cross section 5.5 mb. And, as can be seen from Fig. 3a, our calculations describe well the new data for the thermal cross section with a value of 9.09(10) mb from [25].

Further, the parameterization of the reaction rate in a form of [55] was carried out:

$$N_A\langle\sigma v\rangle = a_1/T_9^{2/3} \cdot \exp(-a_2/T_9^{1/3}) \cdot$$
$$(1.0 + a_3 \cdot T_9^{1/3} + a_4 \cdot T_9^{2/3} + a_5 \cdot T_9 + a_6 \cdot T_9^{4/3} + a_7 \cdot T_9^{5/3}). \tag{18}$$

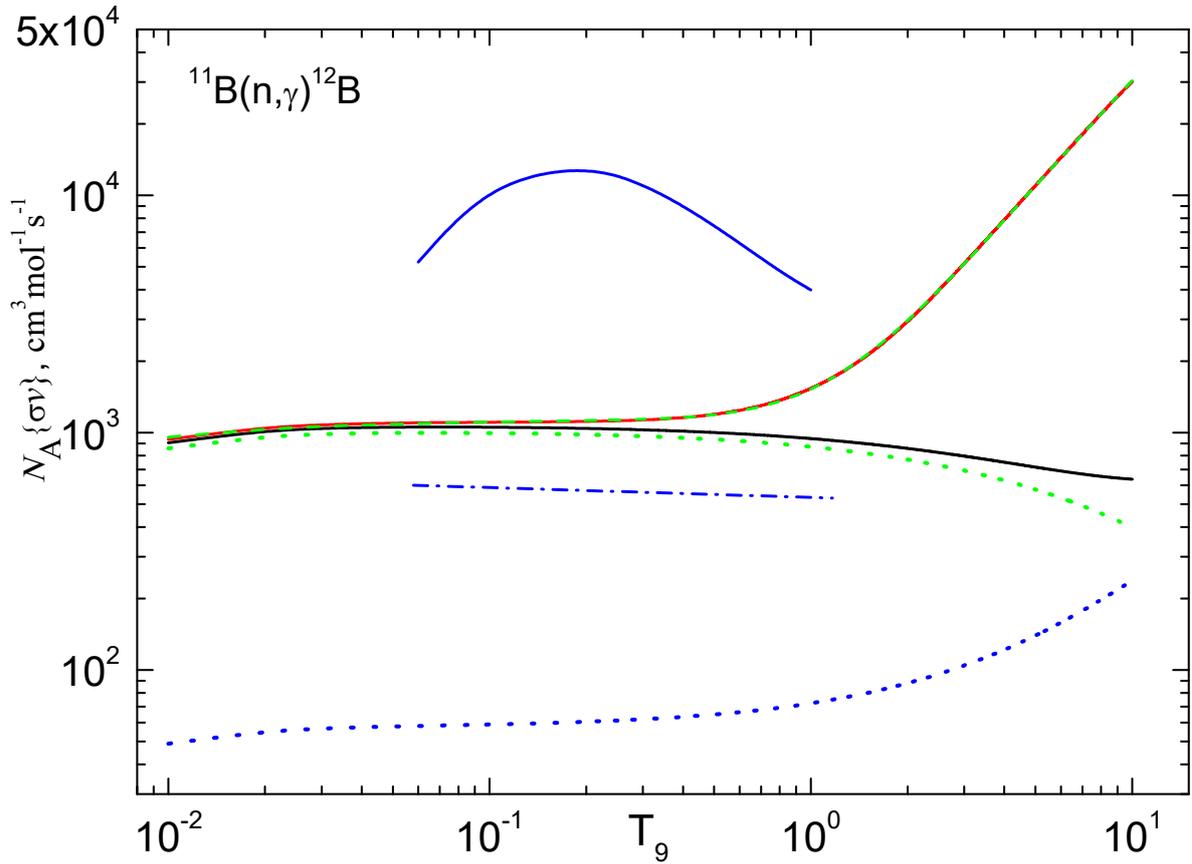

Fig. 4. Reaction rate of the radiative neutron capture on $^{11}$B at low energies. The designations of the curves, as in Fig. 3.

With parameters listed in Table 2.

Table 2. Parameters of analytical parametrization of the reaction rate.

| No. | $a_i$ |
|---|---|
| 1 | 33.91692 |
| 2 | 0.38836 |
| 3 | 55.98557 |
| 4 | -211.765 |
| 5 | 611.5288 |
| 6 | -717.9438 |
| 7 | 327.591 |
| $\chi^2 = 0.01$ | |

The results of that parameterization are shown in Fig. 4 by the green dashed curve, the $\chi^2$ value turns out to be less than 0.01 with 5% errors of the calculated reaction rate.

## 8. Conclusion

So, the results of our calculations for the total cross sections of the $n^{11}$B-capture on the GS are in good agreement with the available experimental data from [25,53] in the energy range of $10^{-5}$ and approximately to $10^2$ keV. The GS potential of $^{12}$B in the $n^{11}$B channel is preliminarily matched with its characteristics, including the asymptotic constant and binding energy, and for the $S$ and $P_0$-scattering potentials, the interactions of the second, third, and fourth excited states of $^{12}$B bound in the $n^{11}$B channel are used.

Taking into account the resonances up to 5 MeV increases the total cross sections in this energy region by approximately two orders of magnitude relative to the results for capture only from the $S$ waves. The reaction rate at a temperature of more than 0.2-0.3 $T_9$ starts to increase and increases about 20 times at 10 $T_9$. As far as we know, such results for the reaction rate were obtained for the first time. They can significantly affect the result of calculations of the yield of $^{12}$B in the primordial nucleosynthesis at high temperatures.

Analysis of the GS potential shows that only in the case of existence the bound FS it is possible to correctly describe available experimental data on total cross sections of the neutron radiative capture on $^{11}$B. Potential without FS, which absolutely coincides with the previous at the description of the characteristics of $^{12}$B, but has other AC sign, overstates on total cross sections to 1.0–1.5 order. These results point out to the necessity of taking into account FSs in interaction potentials of considered particles for the $n^{11}$B system of $^{12}$B.

In conclusion, it should be noted that the new work on the processing of experimental data on the $^{11}$B$(d,p)^{12}$B reaction [52] provides data on the $A_{NC}$ for the four bound ESs of $^{12}$B. This information is promising for further calculations of the characteristics of the synthesis of $^{12}$B in the reaction of radiative neutron capture on $^{11}$B during the capture of the all ESs of the final nucleus. This probably may lead to a revision of certain astrophysical results on the reaction rate.

**Acknowledgements**

This work was supported by the Grant of Ministry of Education and Science of the Republic of Kazakhstan through the program BR05236322 "Study reactions of thermonuclear processes in extragalactic and galactic objects and their subsystems" in the frame of theme "Study of thermonuclear processes in stars and primordial nucleosynthesis" and through the Grant No. AP05130104 entitled "Study of Reactions of the Radiative Capture in Stars and in the Controlled Thermonuclear Fusion" through the Fesenkov Astrophysical Institute of the National Center for Space Research and Technology of the Ministry of Digital Development, Innovation and Aerospace Industry of the Republic of Kazakhstan (RK).